\documentclass[a4paper,11pt]{article}
\usepackage{pos}
\usepackage{sidecap}
\sidecaptionvpos{figure}{t}

\title{Trinity: An Imaging Air Cherenkov Telescope to Search for Ultra-High-Energy Neutrinos}
 \ShortTitle{Trinity: a dedicated IACT to observe UHE neutrinos}

\author*[a]{Anthony M. Brown}
\author[b]{Mahdi Bagheri}
\author[c]{Michele Doro}
\author[b]{Eliza Gazda}
\author[d]{Dave Kieda}
\author[b]{Chaoxian Lin}
\author[b]{Nepomuk Otte}
\author[b]{Ignacio Taboada}
\author[b]{Andrew Wang}

\affiliation[a]{Centre for Advanced Instrumentation, Durham University,
  South Road, Durham, DH1 3LE, UK}

\affiliation[b]{Georgia Institute of Technology, School of Physics \& Center for Relativistic Astrophysics,\\
837 State Street NW, Atlanta, Georgia 30332-0430, USA}

\affiliation[c]{Università di Padova (UniPD), Dipartimento di Fisica e Astronomia (DFA) G. Galilei \\
I-35131 Padova, Italy
}

\affiliation[d]{Department of Physics and Astronomy, University of Utah, 302 Park Building
Salt Lake City, Utah 84112-9016}

\emailAdd{anthony.brown@durham.ac.uk}

\abstract{Earth-skimming neutrinos are those which travel through the Earth’s crust at a shallow angle. For Ultra-High-Energy (E$_{\nu}$ > 1 PeV; UHE) earth-skimming tau neutrinos, there is a high-probability that the tau lepton created by a neutrino-Earth interaction will emerge from the ground before it decays. When this happens, the decaying tau particle initiates an air shower of relativistic sub-atomic particles which emit Cherenkov radiation. To observe this Cherenkov radiation, we propose the \textit{Trinity} Observatory. Using a novel optical structure design, pointing at the horizon, \textit{Trinity} will observe the Cherenkov radiation from upward-going neutrino-induced air showers. Being sensitive to neutrinos in the $1 - 10^4$ PeV energy range, \textit{Trinity}'s expected sensitivity will have a unique role to play filling the gap between the observed astrophysical neutrinos observed by IceCube and the expected sensitivity of radio UHE neutrino detectors.}

\FullConference{37$^{\rm{th}}$ International Cosmic Ray Conference (ICRC 2021)\\
		July 12th -- 23rd, 2021\\
		Online -- Berlin, Germany}

\begin{document}
\maketitle

\section{Introduction}
Neutrino telescopes give us a unique view of the highest energy processes in Universe. The results of the IceCube neutrino telescope over the last decade have accelerated interested in this rapidly evolving field of astronomy. IceCube has observed an all-sky isotropic (extragalactic) diffuse flux of neutrinos \cite{icecube_diffuse}. IceCube has observed a handful of events between 1 PeV and 10 PeV \cite{icecubePeV, glashow}, thus overlapping with Trinity observation band. Standard neutrino oscillations over astrophysical baselines insure a flux of tau neutrinos, to which \textit{Trinity} is sensitive. In fact, the combination of IceCube and Trinity data may prove invaluable in a better determination of the neutrino flux ratio at Earth, which in turn provides information about the nature of the particle accelerators that produce these neutrinos \cite{KW_PRL}. Arguably however, ultrahigh-energy ($E_{\nu}$ > 1 PeV; UHE) neutrinos are an untapped resource when it comes to studying the Universe and addressing some of the most interesting questions in astroparticle physics. 

We know that UHE neutrinos must exist. However, due to the low expected flux of UHE neutrinos, and the low interaction probability of neutrinos in general, the detection of UHE neutrinos represents a formidable challenge. In order to overcome these challenges, a UHE neutrino telescope must observe huge volumes of ice, water, or atmosphere and operate for many years. One instrument that is capable of doing so is \emph{Trinity}. In these proceedings we will outline the \textit{Trinity}'s detection principle, give an overview of \textit{Trinity}'s camera and telescope structure, as well as \textit{Trinity}'s expected sensitivity.

\section{Detection Principle}
\begin{figure}
	\centering\includegraphics[width=1.\linewidth]{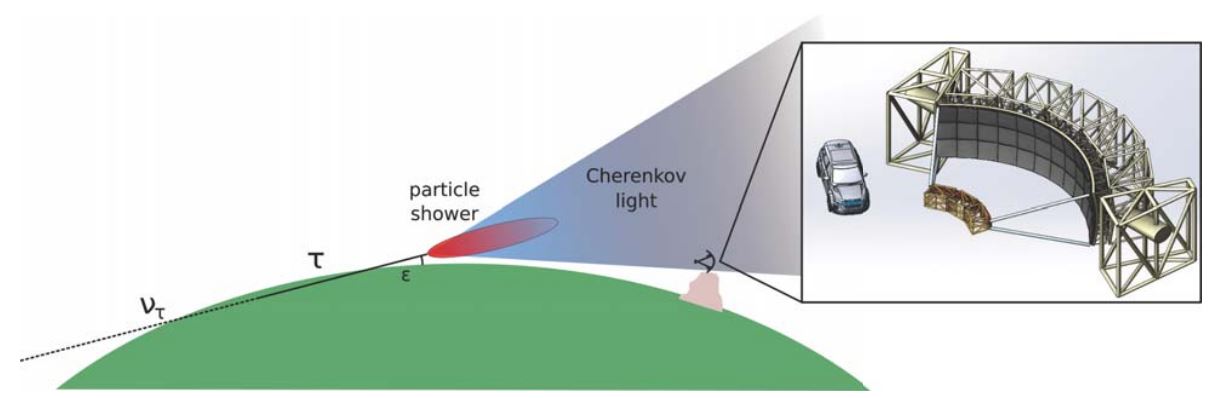}
	\caption{Not-to-scale cartoon of the detection technique employed by \emph{Trinity}. A UHE tau neutrino interacts inside the Earth, resulting in the creation of an energetic tau particle emerging from the ground before decaying and initialising an air shower of billions of Cherenkov radiation emitting particles. \emph{Trinity}’s telescopes observe the Cherenkov radiation and use it to infer the energy of the original neutrino.}
		\label{concept}
\end{figure}

\emph{Trinity} is a dedicated Imaging Air Cherenkov Telescope (IACT) system which detects neutrinos by monitoring the atmosphere and searching for the Cherenkov radiation associated with upward-going air showers. The \emph{Trinity} Observatory will consist of a network of wide field-of-view (FoV) IACT telescopes on mountain tops, which stare at the horizon (see Figure \ref{concept}). In doing so, \textit{Trinity} will employ the so-called `Earth-skimming' technique, a long-proposed and well-established concept which relies on a flux of tau-neutrinos passing through the Earth's crust, close to the surface, and occasionally interacting with the rock. When it does happen, this interaction will create a tau particle which, if energetic enough, will emerge from the ground before decaying. This decay process will create an upwards-going air shower emitting Cherenkov radiation. When viewed with an IACT, the presence and energy of the original tau-neutrino can be inferred from the amount of Cherenkov radiation detected. Importantly, with a wide FoV IACT, a large volume of the atmosphere can be monitored, effectively instrumenting a large fiducial volume. Coupling this potential of the IACT approach with the recent advances in photon-detector technologies – silicon photomultipliers (SiPMs) – and readout electronics means that \emph{Trinity} has the potential to have a competitive sensitivity to UHE neutrinos at modest cost. 
 
\section{Telescope Systems Overview}
\textit{Trinity} will consist of a network of dedicated imaging air Cherenkov telescopes that will observe the horizon searching for these tau neutrino-induced air showers. Using a novel optics design, individual \textit{Trinity} telescopes will have a $60^{\circ}$ wide FoV, a spherical primary mirror, a curved camera focal plane housing $\sim3300$ SiPM pixels and will be sensitive to neutrinos in the $1 - 10^4$ PeV energy range. Here we outline key components of \textit{Trinity}. 
 
\subsection{Camera}
\begin{figure}
	\centering\includegraphics[width=1.\linewidth]{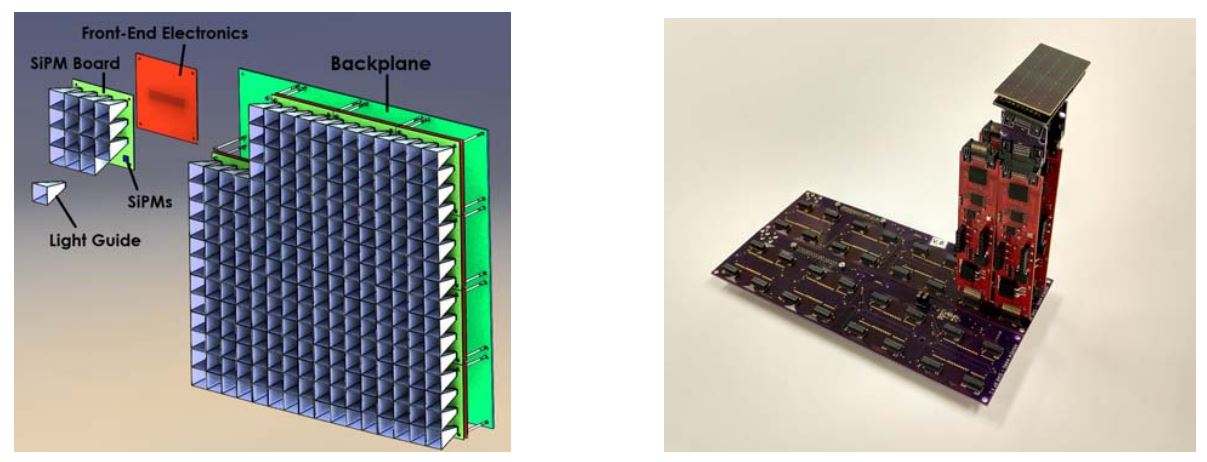}
	\caption{Left: Conceptual drawing of one 256-pixel unit of the camera. A matrix of 4x4 SiPMs coupled to light guides attaches to the front-end electronics board, which forms a module. Sixteen modules connect to one backplane constituting one 256-pixel unit. The 256 signals from one unit connect via micro-coaxial cables to one AGET 256-channel digitizer board. The complete camera comprises 13 units and 13 AGET boards. Right: Partially assembled camera for EUSO-SPB2, which will serve as a blueprint for \textit{Trinity}'s prototype camera.}
	\label{camera}
\end{figure}

The design philosophy of \textit{Trinity}'s camera will be a conceptual copy of the camera being developed for the EUSO-SPB2 balloon mission \cite{euso,eusoicrc2021}. Each camera will have a modular design, consisting of 13 detector modules, with each module having 256 pixels. A CAD drawing of one of these modules can be seen in Figure \ref{camera}.

Each module will house 16 SiPM in a 4x4 matrix. To limit the aberration associated with the spherical optical system design, these SiPMs will be coupled to an array of solid light guides made from PMMA, which will preferentially select Cherenkov light from a small range of incident angles. The electrical signal created by the SiPM when a photon is detected, will first be passed, via micro-coaxial\footnote{The flexible micro-coaxial cables allow for the curvature of the focal plane to be removed.} cables to two 8-channel Multipurpose Integrated Circuit (MUSIC) chips which amplify and then shape the signal from the SiPMs. The MUSIC chip is a low-power Application Specific Integrated Circuit (ASIC) which will provide a 9-bit Digital to Analog Converter (DAC) to adjust the SiPM bias voltage and a current-monitor output for each SiPM channel. The signal is then passed from the MUSIC ASICs to a 256-channel AGET digitizer board. The key features of the AGET board include low power consumption per channel ($< 10$ mW), 100 MS/s sampling rate, 12-bit resolution, 5.12 $\mu$s buffer depth, and negligible deadtime. The AGET board implements the event trigger, and passes all information to the camera backplane. This modular, low-cost, low-power design affords \textit{Trinity}'s camera to be scaleable, which is a necessity to meet the large FoV requirement of the IACT approach. 

\subsection{Telescope Optical Structure}
\begin{figure}
	\centering\includegraphics[width=1.\linewidth]{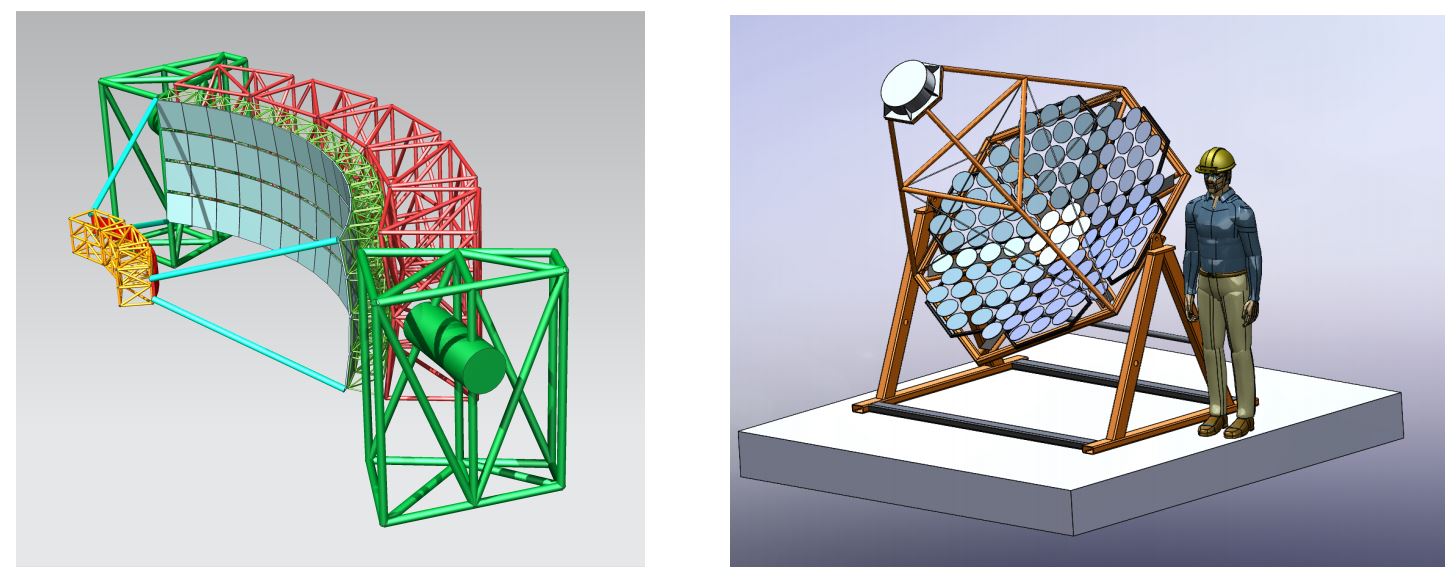}
	\caption{Left: CAD drawing of a $5^{\circ} \times 60^{\circ}$ FoV \textit{Trinity} telescope. Right: CAD drawing of the \textit{Trinity} demonstrator telescope that has recently been funded by the NSF.}
	\label{telescope}
\end{figure}

\textit{Trinity}'s optical system can be seen in Figure \ref{telescope}; the inspiration this design is the spherical geometry proposed for the MACHETE gamma-ray survey telescope \cite{machete}. The primary driver for using the proposed MACHETE optical system is that it possessed an extreme $5^{\circ} \times 60^{\circ}$ FoV, with a angular resolution of $0.3^{\circ}$. 

The optical support structure (OSS) keeps all optics defining elements – mirrors and camera – at their nominal position. \textit{Trinity}'s light-collection surface will consist of 36 tessellated mirror facets, with each facet being one square meter in area. The OSS will rotate in elevation. This elevation control serves two purposes. Firstly, we can point to zenith to verify the telescope’s performance by recording cosmic-ray air showers, and, secondly, the telescope can rotate away from the sun as a redundancy measure to prevent accidental bush fires, should the primary sun-protection of a rolling shutter installed on the OSS to cover the front of the mirrors, fail.

Irrespective of where the telescope points in elevation, the OSS design has to guarantee an optical point spread function that keeps 80\% of the light from a source at infinity within one $0.3^{\circ}$ camera pixel. Considering that existing IACT telescopes, such as VERITAS, MAGIC or H.E.S.S., have demonstrated three times or more stringent PSF requirements, the design of a sufficiently rigid OSS can be based on existing Cherenkov telescopes and is not critical.

\begin{SCfigure}[1.0][!b]
\includegraphics*[angle=0,width=0.55\textwidth]{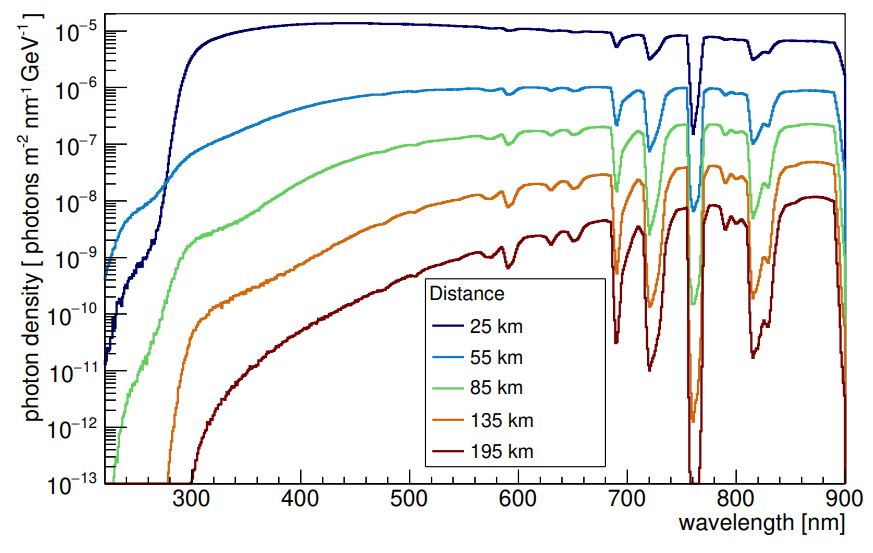}
\caption{Cherenkov photon spectra at the detector for tau-induced air showers starting in distances of 25 km, 55 km, 85 km, 135 km, and 195 km from the detector. At larger distances, the air shower spectrum becomes reddened due to atmospheric transmission properties. These transmission properties need to be accounted for. Taken from \cite{nepomuk}.}
\label{atmosphere}
\end{SCfigure}

\subsection{Calibration systems}
 \textbf{Camera calibration:} During camera operation, to calibrate the data recorded, we will characterise the angular response, pixel gain and linearity. This will be achieved with a compact LED-based light source pulsed light source. Based on a prototype created for the SST component of the Cherenkov Telescope Array (e.g. \cite{brownLED}), we will develop an on-structure calibration system capable of emitting pulses of light at different wavelengths across a wide dynamic range (from 1 to 1000 photoelectrons), with a wide range of pulse duration (4-64 nanoseconds).

\textbf{Atmospheric calibration:} \textit{Trinity} will observe neutrino-induced air showers in the ground-layer of the atmosphere where variations in atmospheric-dust concentration is greatest and most rapid; the effect of these variations can be seen in Figure \ref{atmosphere} . Since \textit{Trinity} will infer the energy of the original tau neutrino by the intensity of the Cherenkov radiation observed, these variations in atmospheric dust content require active monitoring of the atmospheric conditions during observations. To do this, we will employ three complementary methods that are proven or prototyped in other areas of experimental astroparticle physics. These methods are as follows:

\begin{itemize}
    \item \textbf{(i)} monitor the intensity of stars close to the horizon with a separate optical telescope \cite{stars}. 

    \item \textbf{(ii)} observe multi-wavelength light beacons at set distances from the telescope (e.g. \cite{beacons}). 

    \item \textbf{(iii)} cross-check these two calibration methods by periodically flying a UAV-based calibration system, initially designed for CTA (\cite{brown2018,brown2016,jacques2021}), at a range of distances from \textit{Trinity}.
\end{itemize}

\section{Expected Sensitivity}
\begin{figure}
	\centering\includegraphics[width=1.\linewidth]{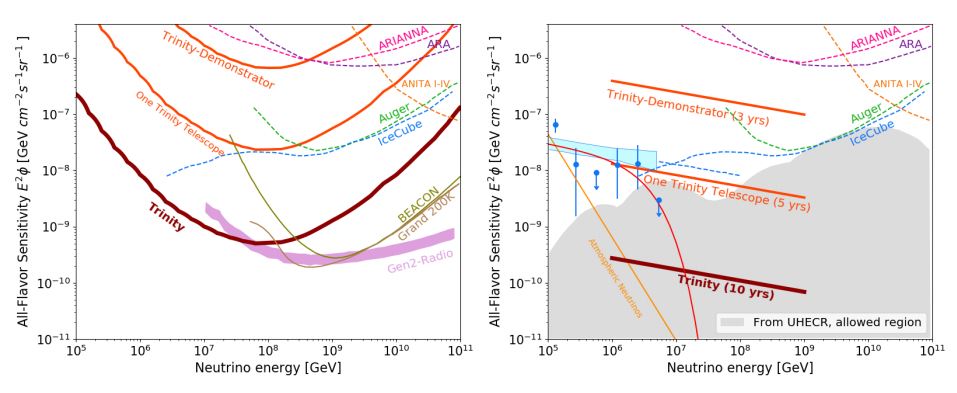}
	\caption{\textit{Trinity}'s differential (left) and integral sensitivity (right). \textit{Trinity} Demonstrator is a single 1\,m$^2$ telescope scheduled to operate for three years. The middle orange curve is for one $5^{\circ} \times 60^{\circ}$ FoV \textit{Trinity} telescope operating for five years. The dark red curve shows the ten-year sensitivity of the complete \textit{Trinity} Observatory. The dashed curves represent published upper limits on the diffuse UHE neutrino flux. The integral sensitivity curves assume a power law with a spectral index of $-2$. Gray shaded is the area of predicted fluxes. The blue data points and the blue shaded bow tie are IceCube's measurements and spectral fit respectively, of the astrophysical neutrino flux. Other experiment sensitivity curves are adapted from \cite{ICG22021}.}
	\label{sensitivity}
\end{figure}

\begin{SCfigure}[1.0][!b]
\includegraphics*[angle=0,width=0.5\textwidth]{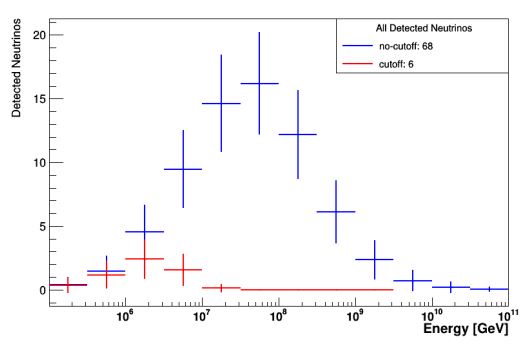}
\caption{Distributions of astrophysical neutrinos, as a function of neutrino energy, detected with \textit{Trinity} over a ten-year long observational, assuming two different spectral extrapolation scenarios. The blue distribution assumes a power-law spectrum for the astrophysical neutrinos without a cutoff. The red distribution assumes the same spectrum but applies an exponential cutoff constrained by the IceCube upper limit at $5$ PeV (see the solid red curve in the right plot of Figure~\ref{sensitivity}).}
\label{number}
\end{SCfigure}

The expected sensitivity for the full \textit{Trinity} Observatory can be seen in Figure \ref{sensitivity}, which also shows the sensitivity of a single $5^{\circ} \times 60^{\circ}$ FoV \textit{Trinity} telescope operating for five years and the smaller \textit{Trinity} demonstrator, which has recently been funded, operating for just 3 years (see Figure \ref{telescope}). A full description of \textit{Trinity}'s sensitivity curves can be found in \cite{wang}. 

From Figure \ref{sensitivity} we can see that one of the unique observing characteristics of \textit{Trinity} is its low-energy threshold, $1$ PeV, compared to other UHE neutrino experiments. This low-energy threshold allows \textit{Trinity} to detect neutrinos in an energy range that overlaps with the astrophysical neutrino spectrum already observed by the IceCube neutrino telescope. This overlap between IceCube and \textit{Trinity} has two important implications: firstly, it ensures that astrophysical neutrinos will be a guaranteed signal for \textit{Trinity} and secondly, raises the possibility of cross-calibrating the performance IceCube and \textit{Trinity}. 

The low-energy sensitivity threshold of \textit{Trinity} also allows us to constrain the high-energy tail of the astrophysical neutrino spectrum observed by IceCube. This ability will be fundamentally important to understanding the origin of the diffuse flux of astrophysical neutrinos observed by IceCube. Figure \ref{number} illustrates \textit{Trinity}'s ability to differentiate between different spectral forms, considering two extreme examples. Assuming a power-law extrapolation of the observed IceCube spectrum up into the UHE regime, \textit{Trinity} will observe 70 neutrino events over a 10-year observing period. Conversely, pessimistically assuming a cut-off to the observed IceCube spectrum, as shown by the red curve in the right plot of Figure \ref{sensitivity}, \textit{Trinity} will still detect 7 neutrino events over a 10-year observing period. Observing any number of neutrino events between these two extremes will be a sensitive probe to existence, or lack, of a spectral cut-off in the diffuse flux of astrophysical neutrinos detected by IceCube. 

\section{Discussion and Outlook}
The observation of UHE neutrinos has the potential to offer some unique insights into long-standing questions in astrophysics and neutrino physics. \textit{Trinity} has a unique role to play in realising this potential. With it low-energy threshold of 1 PeV, \textit{Trinity}'s sensitivity will overlap with the diffuse astrophysical neutrino spectrum already observed by the IceCube. This overlap ensures an astrophysical signal for \textit{Trinity} to observe. 

Regardless of this unique role, \textit{Trinity} is complementary to radio UHE neutrino detectors, which are sensitive to all neutrino flavors. By comparing neutrino fluxes measured with both techniques allows us to probe neutrino physics, such as neutrino oscillations, at the highest energies. The very different detection techniques would also allow us to conduct cross-checks of the systematic uncertainties of the two techniques. 

Given the expected modest costs of \textit{Trinity}, it is feasible to scale the system by deploying several \textit{Trinity} sized detector stations at different locations in both hemispheres. As such, these cross-checks between could potentially take place at the event level in case \textit{Trinity} and an earth-skimming radio detector like GRAND would be deployed at the same site. Furthermore, in the context of multi-sites, we note that modest costs and scalability of \textit{Trinity} also affords us the possibility of increasing the sensitivity of the \textit{Trinity} Observatory by adding further telescope sites.  

Finally we note that funding has recently be awarded for the creation of the \textit{Trinity} demonstrator prototype system showing in Figure \ref{telescope} \& \ref{sensitivity}. We will install the prototype telescope near Frisco Peak's summit (9664 ft. above sea level). This site has the benefit of infrastructure already in place for previous experiments, and also has surrounding mountains on which calibration light beacons can be placed. 


%
%
%


\begin{thebibliography}{99}
\bibitem{icecubePeV} Aarsten, M.G., et al., 2013, PRL, 111, 1103
\bibitem{icecube_diffuse} Aarsten, M.G., et al., 2014, PRL, 113, 1101 
\bibitem{eusoicrc2021} Bagheri, M., et al., 2021, Proceedings of the 37th International Cosmic Ray Conference
\bibitem{brown2018} Brown, A.M., 2018, APh, 97, 69
\bibitem{brown2016} Brown, A., et al., 2016, \textit{Ground-based and Airborne Telescopes VI, volume 9906 of Society of Photo-Optical Instrumentation Engineers (SPIE) Conference Series}, page 99061W
\bibitem{brownLED} Brown, A.M. et al., Proceedings of the 34th International Cosmic Ray Conference, July 2015
\bibitem{machete} Cortina, J., Lopez-Coto, R. and Moralejo, A., 2019, APh, 72, 46
\bibitem{stars} Ebr, J., et al., 2021, AJ, 162, 6
\bibitem{glashow} Glashow, S.L., 1960, Phys. Rev. 118, 316
\bibitem{ICG22021} IceCube-Gen2 Collaboration, 2021, Journal of Physics G: Nuclear and Particle Physics, 48, 6 
\bibitem{KW_PRL} Kashti, T. \& Waxman, E., 2005, PRL, 95, 181101
\bibitem{jacques2021} Muller, J., Brown, A.M. \& de Naurois, M., 2021, Proceedings of the 37th International Cosmic Ray Conference
\bibitem{euso} Otte, A.N., et al., 2019, Proceedings of the 36th International Cosmic Ray Conference
\bibitem{nepomuk} Otte, A.N., 2019, PRD, 99, 038012
\bibitem{wang} Wang, R., et al., 2021, Proceedings of the 37th International Cosmic Ray Conference 
\bibitem{beacons} Wiencke, L.R., et al., 1999, NIM A, 428, 593


\end{thebibliography}
\end{document}